\documentclass[english,letterpaper,twocolumn,showpacs,pra,aps]{revtex4}
\usepackage{graphicx}
\usepackage{amssymb}

\makeatletter

\large  

\input epsf

\makeatother

\usepackage{babel}
\makeatother
\begin{document}

\title{Casimir interaction of arbitrarily shaped conductors}

\author{Joseph P. Straley$^{1}$ and Eugene B. Kolomeisky$^{2}$}

\affiliation
{$^{1}$Department of Physics and Astronomy, University of Kentucky,
Lexington, Kentucky 40506-0055, USA\\
$^{2}$Department of Physics, University of Virginia, P. O. Box 400714,
Charlottesville, Virginia 22904-4714, USA}

\begin{abstract}
We review a systematic practical implementation of the multiple scattering formalism due to  Balian and Duplantier [R. Balian and B. Duplantier, Ann. Phys. (NY) \textbf{104}, 300 (1977); \textbf{112}, 165 (1978)] for the calculation of the Casimir interaction between arbitrarily shaped smooth conductors.  The leading two-point scattering term  of the expansion has a simple compact form, amenable to exact or accurate numerical evaluation.  It is a general expression which improves upon the proximity force and pairwise summation approximations.  We show that for many  geometries it captures the bulk of the interaction effect.  The inclusion of terms beyond the two-point approximation provides an accuracy check and explains screening.  As an illustration of the power and versatility of the method we re-evaluate sphere-sphere and sphere-plane interactions and compared the results with previous findings that employed different methods.  We also compute for the first time interaction of a hyperboloid (mimicking an atomic force microscope tip) and a plane.  We also analyze the anomalous situations involving long cylindrical conductors where the two-point scattering approximation fails qualitatively. In such cases analytic summation of the entire scattering series is carried out and a topological argument is put forward as an explanation of the result.    We give the extension of this theory to the case of finite temperatures where the  two-point scattering approximation result has a simple compact form, also amenable to exact or accurate numerical evaluation.                 
\end{abstract}

\pacs{03.70.+k, 11.10.-z, 11.10.Gh, 42.50.Pq}

\maketitle

\section{Motivation}
 
Casimir interactions are the macroscopic response of the physical vacuum to the introduction of material objects. They were first derived as an attractive force between perfectly conductive parallel plates induced by the zero-point motion of 
the electromagnetic field \cite{Casimir}.  There is convincing experimental evidence for the reality of these forces \cite{experiment} and a vast body of literature dedicated to various aspects of the phenomenon \cite{CasReviews}. 

The goal of this topical review is to provide a practical guide to the computation of the Casimir interactions between arbitrarily shaped  smooth conductors.  Since the introduction of a conductor  into a previously empty space modifies the electromagnetic spectrum, 
and thus the zero-point energy, the Casimir energy can be heuristically presented in the form \cite{Casimir}
\begin{equation}
\label{casimirnumber}
{\mathcal E}  =  \frac {\hbar}{2} \sum_{\alpha} (\omega_{\alpha} - {\bar \omega}_{\alpha})
\end{equation}
where $\bar \omega_{\alpha}$ and $\omega_{\alpha}$ are the mode frequencies before and after the introduction of the conductor.  This idea can be made rigorous by including a smooth cutoff at large frequency; a large variety of cutoff functions yield the same result \cite{KZLS}.  The case of \textit{perfect} conductors which is our focus plays a special role because Casimir forces are cutoff-independent thus only reflecting the geometry of the problem \cite{BD}.   

When the conductor is a collection of several disjoined objects, part of the Casimir energy depends on the relative location and orientation of the objects; this is the interaction energy $\mathcal{E}_{int}$.  Although it can in principle be calculated to arbitary accuracy \cite{BD,exact,EGJK,johnson}, in practice such calculations encounter significant computational difficulties (the parallel plates geometry is the only exception). Then it is useful to have approximate methods of computation, especially when there is a systematic means to improve the result.  

Casimir \cite{Casimir} calculated the force between parallel planes.  A simple method to study non-planar geometries, the Derjaguin or proximity force approximation (PFA) extends this to nonplanar objects by treating them as a superposition of infinitesimally small facing parallel plane segments \cite{Derjaguin}.  It is assumed that only the facing sides of the interacting objects need be considered. The method is restricted to the case that the distance of closest approach is significantly smaller than the objects themselves.

A complementary and equally simple method (proposed by Mostepanenko and Sokolov \cite{Most}) is based on the interaction between small objects at large separation given by Casimir and Polder \cite{CP}, and extending this to intermediate sizes and distances by regarding macroscopic objects as a collection of infinitesimally small volume elements and performing a pairwise summation of elementary two-body interactions. 

Even though in many geometries these approximation schemes give reasonably good results, they suffer from two fundamental shortcomings:  (i) they assume additivity of Casimir interactions, though this is known not to be the case; (ii) in their original form they are uncontrolled, thus making it difficult to independently judge and improve the results.  

In recent years significant progress has been made in computing analytic corrections to the PFA which is necessary for interpretation of precise modern day measurements of the Casimir force.  For example, the leading order proximity corrections (relevant to the case that the distance of closest approach is much smaller than object sizes)  for the interaction between a cylindrical conductor that is parallel to a conductive plane \cite{Bordag06} or for a spherical conductor facing a conductive plane \cite{Teo} have been computed.  These findings nicely fit with the idea of an approximation scheme in which the Casimir result for parallel plates is supplemented by a phenomenological expansion in the transverse derivatives of the height function \cite{Bimonte}.  The approach advocated in this review complements these studies because it is not limited by the proximity approximation or shape and gives usable and improvable approximation at all separation distances of the objects.                 

This review is based on the multiple scattering expansion technique due to Balian and Duplantier (BD) \cite{BD}.  We will present a systematic approach to the calculation of Casimir interactions between arbitrarily shaped conductors, which overcomes  the shortcomings of previous treatments \cite{Derjaguin,Most}.  The interaction energy is represented as a sum of terms; the leading term is a two-point integral, with one point on each surface.  We show that for many geometries and arbitrary separation of the conductors this term ($\mathcal{E}_{2}$ in what follows) captures the bulk of the interaction energy $\mathcal{E}_{int}$, while being as simple and compact as previous approximations \cite{Derjaguin,Most}.  For these geometries the Casimir interactions can be viewed to be approximately additive.  Including the four-point scattering contribution $\mathcal{E}_{4}$ gives a useful check on the accuracy: usually it is small, but in some problem cases it is not.   

However, there are geometries where the result supplied by $\mathcal{E}_{2}$ is qualitatively wrong in the large separation distance limit;  in such cases the entire multiple scattering series is required to obtain the right answer for $\mathcal{E}_{int}$;  we put forward a topological argument which purports to explain such anomalous cases.        

Pertinent to the situations when $\mathcal{E}_{2}$ captures the bulk of the effect, we give finite-temperature generalization of the result, i.e. the expression for the Casimir interaction free energy $\mathcal{F}_{2}$ which, like $\mathcal{E}_{2}$, has a simple compact form.       

Despite its generality, as computational technology the BD method has received very little attention in the literature even though a prospect of using it to calculate Casimir interactions between arbitrarily shaped conductors has been mentioned \cite{EGJK}.  A 2010 review of the subject even stated that the BD  "method has not proved workable in practice" \cite{French}.  This is possibly due to complexity of multiple surface integrals involved into evaluation of subsequent terms of the multiple scattering expansion.  However these integrals are numerically tractable.  Specifically we will use Monte Carlo integration which allowed us to implement the BD method in the form of a flexible practical tool amenable for calculation of Casimir interaction between arbitrarily shaped conductors (finite element methods would undoubtedly be more efficient and precise, but less adaptable).  We also note that the BD method was recently utilized by us to solve a related problem of computation of the Casimir \textit{self-energy} of an arbitrarily-shaped solitary smooth conductor \cite{SK}.  There is no doubt that the BD method can be also used to compute $3$- and generally $N$-body Casimir interactions of arbitrarily shaped conductors.   

The rest of this review is organized as follows:  

In Section II, following BD \cite{BD}, we give general expression for the Casimir energy $\mathcal{E}$ as a multiple scattering series.  

The part of that energy $\mathcal{E}_{int}$ that depends upon the relative position and orientation of the conductors is discussed in Section III where we observe (Section IIIA) that the leading two-point contribution of the BD expansion $\mathcal{E}_{2}$ has a simple compact form of a double surface integral over the two surfaces of the conductors.  This expression is nearly equivalent to a representation of the Casimir interaction as a superposition of elementary Casimir-Polder \cite{CP} interactions, which has been  proposed on phenomenological grounds \cite{Most}.  For simple geometries $\mathcal{E}_{2}$ can be computed analytically and several examples of successful calculations (two small conductors, small conductor and a plane, two parallel planes, and two spheres) are given.  However, there is a qualitative failure of the two-point approximation for geometries involving long cylinders.  The four-point interaction term is given explicitly in Section IIIB while higher-order contributions are discussed in Section IIIC.  What sets the BD method apart from majority of other approaches (outlined in Section IIID) is that it does not rely on separability of Maxwell's equations which is limited to  a few high symmetry cases.  The only other method known to us where separability is not essential is that due to Johnson and collaborators \cite{johnson} which we view as complementary to the BD approach.

Applications of the BD expansion to a series of concrete problems are given in Section IV.  We begin with the parallel plane geometry (Section IVA) followed by the plane-arbitrary surface geometry (Section IVB) where we observe that the two-point term of the BD expansion derives (up to numerical factor close to unity) the PFA;  a counterpart of this result for the sphere-arbitrary surface geometry is also given.  We then re-evaluate the Casimir interaction in the sphere-plane (Section IVC) and sphere-sphere (Section IVD)  geometries and demonstrate that the two-point approximation captures about $90\%$ of the result found in the literature while the numerically evaluated four-point term accounts for most of the discrepancy.  As a potentially important practical application, in Section IVE we study the geometry of hyperboloid of rotation (mimicking a rounded tip of atomic force microscope) facing a plane, and investigate dependences of the Casimir interaction on the plane-tip separation and tip's opening angle.  As before, the two-point term is computed analytically while the four-point contribution is evaluated numerically.  The non-additivity of the Casimir effect is discussed in the context of an object near a conducting slab of finite thickness.  Here the physical effect of screening plays an important role which the two-point approximation misses.  By analyzing the sphere-slab geometry (Section IVF) beyond the two-point approximation we explicitly demonstrate that the screening mechanism is built into the BD expansion;  simultaneously this provides an explanation of the success of the PFA for closely positioned conductors.  In Section IVG we focus on anomalous geometries involving long cylinders where the two-point approximation fails qualitatively and the entire BD series needs to be accounted for.  We relate this difficulty to change of space topology due to the presence of cylindrical conductor, and then, for the plane-parallel cylinder geometry, carry out analytic summation of the entire multiple scattering series, and confirm the result found in the literature.

We conclude (Section V) by deriving an expression for the Casimir interaction free energy in the two-point approximation, $\mathcal{F}_{2}$, which has a simple and compact form.  It encompasses the classical (high-temperature limit) and is expected to exhibit the same level of accuracy as its zero-temperature counterpart $\mathcal{E}_{2}$.  

\section{Casimir energy}

BD \cite{BD} have given an approach to the Casimir problem that allows calculation of $\mathcal E$ for a conducting surface of arbitrary shape \cite{SK}.  The starting point of BD's method is the observation that the electromagnetic response (fields and induced currents) to a current source perturbation (such as an oscillating magnetic dipole) will diverge at the resonant frequencies.  Viewed as analytic functions of the frequency, the corresponding Green functions have poles along the real frequency axis; these can be used to construct a contour integral representation for $\mathcal E$.  This is the basis of many approaches to the calculation of the Casimir energy \cite{contour,BG}.    

The magnetic field due to a perturbing current source on one side of the surface is completely screened so that there is no response to the perturbing current on the other. The screening is accomplished by a current ${\vec j}_{s} (\vec r)$ on the surface.  BD observe that the surface current is related by Ampere's law to the tangential components of the total field, which is due both to the perturbing current and to the surface current elsewhere on the surface.  This leads to the integral equation
\begin{equation}
\label{BD1}
\vec j ({\vec r}_{2}) = {\vec j}_{s} ({\vec r}_{2})+ \int {\textbf K}({\vec r}_{2}, {\vec r}_{1}) \cdot \vec j ({\vec r}_{1}) d^{2}r_{1}
\end{equation}
where the integral is over the surface and $\textbf K$ is the tensor that describes how a current $\vec j ({\vec r}_{1})$ at ${\vec r}_{1}$ would induce a current $\vec j ({\vec r}_{2})$ at ${\vec r}_{2}$, ignoring all the other currents, so that the electromagnetic fields propagate as if in empty space.  The integral equation will give divergent response at the mode frequencies.  BD show how to use this property to calculate the change in the density of states caused by the introduction of the conductors.  From this the Casimir energy can be calculated as a contour integral which is most conveniently evaluated by integrating along the imaginary wave number axis $y=\omega/ic$.  Here the tensor $\textbf K$ is   
\begin{equation}
\label{BD2}
{\textbf K}(\vec r_{2}, \vec r_{1})
= - \frac {e^{-y\rho}(1+y\rho)}{2\pi \rho^{2}} (\hat \rho \otimes \hat n_{2} - {\textbf I} \hat \rho \cdot \hat n_{2})
\end{equation}
where $\vec \rho = \vec r_{2} - \vec r_{1}$, $\rho = |\vec \rho|$,  $\hat \rho = \vec \rho/\rho$, $\otimes$ denotes the dyadic product, $\hat n_{2}$ is the outward normal to the surface at $\vec r_{2}$, and $\textbf{I}$ is the unit tensor.  The expression for the Casimir energy is

\begin{equation}
\label{casimir1}
{\cal E} = \frac{\hbar c}{\pi} \int_0^{\infty} \left (\Phi(iy) - \Phi(i\infty)\right ) dy
\end{equation}
where
\begin{equation}
\label{casimir2}
\Phi(iy)= \sum_{even ~m\geqslant 2} \Phi_{m}(iy)
\end{equation}
The functions $\Phi_{m}$  are given by
\begin{equation}
\label{casimir3}
\Phi_{m} = \frac {y}{2m} \frac {d}{dy} \text {Tr} {\textbf K}(\vec r_{m},\vec r_{m-1})...{\textbf K}(\vec r_{2},\vec r_{1}){\textbf K}(\vec r_{1},\vec r_{m})
\end{equation}
where the operation $\textrm{Tr}$ includes the integration of the variables $\vec r_{i}$ over all surfaces, as well as the trace of the product of the tensors ${\textbf K}(\vec r_{i},\vec r_{i-1})$.   The terms can be represented as closed paths that visit $m$ sites on the surfaces.  The paths having an odd number of points make no contribution because reversing the path changes the sign of each factor $\textbf K$, and both paths occur within the integral over paths.

The sum in Eq.(\ref{casimir2}) can be done, giving
\begin{equation}
\label{casimirx}
\Phi(iy) = - y\frac{d}{dy} Tr \ln (1- {\textbf K}\cdot {\textbf K} )
\end{equation}
This is the starting point for many calculations of the Casimir interaction energy \cite{exact,EGJK}. 
 
\section{Casimir interaction}

We have shown elsewhere \cite{SK} how to use this result to calculate the self-energy of a single conducting object. It applies equally well to the calculation of the energy of interaction of two objects:  each surface integral is over the union of the surfaces of the two objects, and can be expanded into the $2^{m}$ assignments of the points of the paths to one surface or another.  

Several important simplifications result.   The first is that we can drop the paths that are entirely confined to one surface or another, because these give the self-interaction of one of the two objects and is not an interaction between them. Although there are relatively few of these paths, they are the most singular, because they can be arbitarily short; removing them guarantees a finite path length and thus removes the need for the subtraction $-\Phi(i\infty)$ in Eq.(\ref{casimir1}).  This in turn allows an integration by parts in Eq.(\ref{casimir1}), so that the expression for the Casimir interaction reduces to 
${\mathcal E}_{int} = \sum_{m \ge 1} {\cal E}_{2m}$, with
\begin{equation}
\label{casimir4}
{\mathcal E}_{2m} = - \hbar c\int_{0}^{\infty} \frac {dy}{2\pi m}\textrm{Tr} 
{\textbf K}(\vec r_{m},\vec r_{m-1})...{\textbf K}(\vec r_{2},\vec r_{1}){\textbf K}(\vec r_{1},\vec r_{m})
\end{equation}
where the $\textrm{Tr}$ again includes integrals of each of the $\vec r_{i}$ over all surfaces, but excluding the cases where all points are on the same surface.

\subsection{Two-point term}

The final simplification is that for small $m$ it is practicable to do the $y$ integration in Eq.(\ref{casimir4}) first.  For the two-point contribution the only relevant paths go from surface of conductor $A$ to that of conductor $B$ and back again, with the result for the two-point function
\begin{equation}
\label{2_point}
\Phi_{2}(iy)=-\frac{y^{2}}{4\pi^2} \int \frac{(\vec \rho \cdot d\vec S_{A})(\vec \rho \cdot d\vec S_{B})}{\rho}\frac{d}{d\rho}\left (\frac{e^{-2y\rho}}{\rho^{2}}\right )
\end{equation}
where $\vec \rho = \vec r_{A} - \vec r_{B}$, $\vec r_{A}$ ($\vec r_{B}$) is the point on surface $A$ ($B$) and $d\vec S_{A}$ ($d\vec S_{B}$) is the vector surface element at $\vec r_{A}$ ($\vec r_{B}$).  
The interaction energy is then given by
\begin{equation}
\label{int1}
{\mathcal E}_{2} = \frac {5\hbar c}{16\pi^3} \int
\frac {(\vec \rho \cdot d\vec S_{A})(\vec \rho \cdot d\vec S_{B})}
{\rho^{7}}
\end{equation}
The interaction (\ref{int1}) depends on the relative orientation of the surface elements;  specifically anti-parallel surface elements attract while parallel elements repel each other.  

Application of Gauss's theorem to Eq.(\ref{int1}) once and then once again gives two further representations of the result
\begin{equation}
\label{int2}
{\mathcal E}_{2} = -\frac {15\hbar c}{16\pi^{3}} \int 
\frac {(\vec \rho \cdot d\vec S_{B})dv_{A}}{\rho^{7}}
\end{equation}
\begin{equation}
\label{int3}
\mathcal{E}_{2}=- \frac {15\hbar c}{4\pi^{3}} \int \frac{dv_{A} dv_{B}}{\rho^{7}} 
\end{equation}
where $dv_{A}$ ($dv_{B}$) is the volume element for conductor $A$ ($B$) and the corresponding integrals are over the interior of one conductor or the other.  Eq.(\ref{int3}) may be taken as an indicator of approximate additivity of Casimir-Polder interactions \cite{CP}.  This was assumed by Mostepanenko and Sokolov \cite{Most} who proposed the functional form (\ref{int3}) to complement the PFA.  We see that the BD method justifies the conjecture \cite{Most} and derives an amplitude;  including higher order terms allows for systematic improvement on Eqs.(\ref{int1}-\ref{int3}). 

Eq.(\ref{int3}) implies that interaction is necessarily attractive (to this order of approximation) for any pair of objects, so that a repulsive Casimir interaction will require that the higher order terms play a dominant role. Eqs.(\ref{int1} - \ref{int3}) are equivalent representations of the first term of an expansion which we believe to be exact.  As we shall see, Eq.(\ref{int1}) is a good approximation for many standard problems.   

If the distance $D$ between the two conductors significantly exceeds their linear sizes, the interaction energy (\ref{int3}) can be written explicitly as
\begin{equation}
\label{2_separate_conductors}
\mathcal{E}_{2}= - \frac{15\hbar c v_{A}v_{B}}{4\pi^{3}D^{7}}
\end{equation}   
which agrees with the functional form of the Casimir-Polder original result \cite{CP}.  For two spherical conductors of radii $A$ and $B$ we find $\mathcal{E}_{2}= - 20 \hbar cA^{3}B^{3}/(3\pi D^{7})$.  The exact result in this case \cite{Feinberg,Boyer}, $\mathcal{E}_{int}= - 143\hbar c A^{3}B^{3}/(16\pi D^{7})$ has about $25\%$ larger amplitude.

Similarly, for interaction energy of conductor $A$ placed a large distance $D$ away from conductive plane, Eq.(\ref{int2}) predicts
\begin{equation}
\label{plane_conductor}
\mathcal{E}_{2}=-\frac{3\hbar cv_{A}}{8\pi^{2}D^{4}}
\end{equation}
This again agrees with the functional form of the Casimir-Polder result \cite{CP}.  For spherical conductor of radius $A$ we find $\mathcal{E}_{2}=-\hbar cA^{3}/2\pi D^{4}$.  The exact result in this case \cite{Boyer}, $\mathcal{E}_{int}=-9\hbar cA^{3}/16\pi D^{4}$, is larger in magnitude by a factor of $9/8$.   

According to Eq.(\ref{int3}), the two-point approximation to the Casimir energy can be interpreted in terms of a pairwise scalar interaction between uniform density clouds of "Casimir charge" that fill the two objects.   By doing the integral over one of the objects we can define a "Casimir potential" ${\mathcal V}_{B}(\vec r)$ due to conductor $B$ such that the two-point contribution to the energy is given by the integral of the potential over the interior of the second object $A$:
\begin{equation}
\label{poten0}
{\mathcal E}_{2} = \int {\mathcal V}_{B}(\vec r) dv_{A}
\end{equation}

For simple geometries this can be used to find closed form of the interaction energy at arbitrary separation of the conductors.  For the plane, the potential depends only on the distance $z$ from it, in the form
\begin{equation}
\label{plane}
{\mathcal V}_{plane}(z) = - \frac {3\hbar c}{8\pi^{2} z^{4}}
\end{equation}

$\bullet$ Integrating this over the half-space $z>D$ gives the interaction per unit area of parallel planes.  To this order of approximation
\begin{equation}
\label{||_planes}
\frac{\mathcal{E}_{2}}{\mathcal{A}}=-\frac{\hbar c}{8\pi^{2}D^{3}}
\end{equation}
which differs from Casimir's result \cite{Casimir} ${\mathcal E}_{int}/{\cal A} = - \pi^{2}\hbar c/(720 D^{3})$ by a factor $\zeta(4) = \pi^{4}/90=1.0824$  (BD \cite{BD} show how the higher $m$ contributions ${\mathcal E}_{m}$ give rise to this correction).

$\bullet$ Integrating the potential (\ref{plane}) over a sphere of radius $A$ at distance $D$ from the plane gives the interaction energy for sphere and plane
\begin{equation}
\label{sphere_plane}
{\mathcal E}_{2} = - \frac{\hbar cA^{3}}{2\pi(D^{2}-A^{2})^{2}}
\end{equation}
For $D\gg A$ this agrees with Eq.(\ref{plane_conductor}) while for $D \approx A$ it becomes \cite{BD}
\begin{equation}
\label{sphere_pfa+correction}
\mathcal{E}_{2}\approx -\frac{\hbar c A}{8\pi X^{2}} +\frac{\hbar c}{8\pi X}=-\frac{\hbar c A}{8\pi X^{2}}\left (1-\frac{X}{A}\right )
\end{equation}
where $X=D-A$ is the distance of closest approach.  The first term in (\ref{sphere_pfa+correction}) is very close to the PFA result (again too small by the factor $\zeta(4)$) while the second represents a correction to the PFA in terms of the dimensionless proximity parameter $X/A$.  It compares reasonably well with the result of an asymptotically exact calculation of the sphere-plane interaction energy $\mathcal{E}_{int}(X\ll A)\propto (1-1.69X/A)$ \cite{Teo} especially given that Eq.(\ref{sphere_plane}) is just the two-point result.  Below we will show that it also works reasonably well at all separations.  The same functional form as in Eq.(\ref{sphere_plane}) (with a different amplitude of phenomenological origin) was given previously \cite{Most}.

For a sphere of radius $A$, the potential depends on the distance $r$ from its center
\begin{equation}
\label{spherepot}
{\mathcal V_{sphere}(r)} = - \frac {\hbar cA^{3}}{\pi^{2} r} \frac {5r^{2} + A^{2}} {(r^{2}-A^{2})^{4}}
\end{equation}
For $r\approx A$ this reduces to the potential of the plane (\ref{plane}).  The interaction energy with a second sphere of radius $B$ and center at $D$ is
\begin{eqnarray}
\label{twospheres}
{\cal E}_{2}=-\frac{\hbar c}{8\pi D}&\Bigg\{&
\ln \left (\frac {R_{12} R_{21}}{R_{11} R_{22}}\right )-\frac {2D^{2}}{R_{11}
R_{22}}
+ \frac {2D^{2}}{R_{12}R_{21}}\nonumber\\
&+& \frac{AB}{R_{11}^{2}}
+\frac{AB}{R_{12}^{2}}
+\frac{AB}{R_{21}^{2}}
+\frac{AB}{R_{22}^{2}} \Bigg\}
\end{eqnarray}
where 
$R_{11} = D+A+B$, $R_{12} = D+A-B$, $R_{21}=D-A+B$, and
$R_{22}=D-A-B$.  For large $D$ this reduces to Eq.(\ref{2_separate_conductors}).  For 
$D$ close to $A+B$, we find ${\mathcal E}_{2} \approx -\hbar c AB/[8\pi (A+B)(D-A-B)^{2}]$.

Other geometries that do not allow explicit analytic calculation of $\mathcal{E}_{2}$ can be studied by numerical evaluation of Eqs.(\ref{int1}), (\ref{int2}), or (\ref{int3}).  We observe that when the separations between objects is large, Eq.(\ref{int1}) is less appropriate because the near and far sides of each object have contributions of opposite sign; when the objects are near each other, however, Eq.(\ref{int1}) may be more efficient.

$\bullet$ These encouraging results are offset by geometries involving macroscopically long cylindrical conductor.  For example, for a cylinder of length $\mathcal{L}$ and radius $A$ at distance $D\geqslant A$ from conductive plane, we find for an interaction energy per unit length  
\begin{equation}
\label{cylinder_plane}
\frac{\mathcal{E}_{2}}{\mathcal{L}}=-\frac{3\hbar cA^{2}D}{8\pi (D^{2}-A^{2})^{5/2}}
\end{equation}
The same functional form (with different amplitude) was given previously \cite{Most}.  For $D\gg A$ this agrees with Eq.(\ref{plane_conductor}) while for $D \approx A$ it becomes
\begin{equation}
\label{cylinder_pfa+correction}
\mathcal{E}_{2}\approx-\frac{3\hbar c A^{1/2}}{32\sqrt{2}\pi X^{5/2}}\left (1-\frac{X}{4A}\right )
\end{equation}
where the leading order term is very close to the PFA result \cite{Emig_cylinder} ($\mathcal{E}_{2}/\mathcal{E}_{PFA}=\zeta^{-1}(4)=90/\pi^{4}=0.9239$) while the PFA plus proximity correction, $\mathcal{E}_{2}\propto (1-X/4A)$, compares reasonably well with asymptotically exact result, $\mathcal{E}_{int}\propto (1-0.48103 X/A)$ \cite{Bordag06}.  However, at intermediate cylinder-plane separations $D\gg A$ (where $\mathcal{L}$ is the largest length scale in the problem), both Eq.(\ref{cylinder_plane}) and the PFA are qualitatively different from the result due to Emig et al.  \cite{Emig_cylinder}, $\mathcal{E}_{int}/\mathcal{L} \simeq -\hbar c/(D^{2} \ln(D/A))$. 

Similarly, for the Casimir interaction of two well-separated cylinders of arbitrary cross-sectional areas $\sigma_{A}$ and $\sigma_{B}$ we find
\begin{equation}
\label{cylinder_cylinder}
\frac{\mathcal{E}_{2}}{\mathcal{L}}=-\frac{4\hbar c\sigma_{A}\sigma_{B}}{\pi^{3}D^{6}}
\end{equation}
which must be valid if $D$ is the largest length scale of the problem.  However, this disagrees with the asymptotically exact result due to Rahi et al. \cite{Rahi} for cylinders of radii $A$ and $B$,  $\mathcal{E}_{int}/\mathcal{L}\simeq -\hbar c/[D^{2}\ln(D/A)\ln(D/B)]$ valid when $D \gg A,B$ and $\mathcal{L}$ is the largest length scale of the problem.  These anomalous cases will be further discussed below.  

\subsection{Four-point interaction}

The dependence of $\textbf K$, Eq.(\ref{BD2}), on the imaginary wavevector $y$ is entirely contained in the scalar prefactor, so that the integral over $y$ in Eq.(\ref{casimir4}) can always be done first.  For the four-point term ${\mathcal E}_{4}$ the integrand is
\begin{eqnarray}
\label{yint4}
J &=& \int_{0}^{\infty}  (1+y\rho_{43})(1+y\rho_{32})(1+y\rho_{21})\nonumber\\
&\times& (1+y\rho_{14})\exp(-y[\rho])dy
\end{eqnarray}
where the abbreviation
$
[\rho^{n}] = \rho_{43}^n + \rho_{32}^n+\rho_{21}^n+\rho_{14}^{n}
$
has been introduced.  A little algebra shows that
\begin{equation}
\label{yint2}
J = 3[\rho]^{-1} - [\rho]^{-3} [\rho^{2}] + \frac{6 [\rho][1/\rho] + 24}{[\rho]^{5}}\Pi
\end{equation}
in which we have introduced the further abbreviation
$\Pi =  \rho_{43}\rho_{32}\rho_{21}\rho_{14}$.
This reduces the problem to the evaluation of
\begin{equation}
\label{fourpt2}
{\cal E}_{4} = -\frac {\hbar c}{4}\textrm{Tr} \frac{J}{(2\pi)^{5} \Pi^{2}} {\textbf k}(4,3){\textbf k}(3,2) {\textbf k}(2,1){\textbf k}(1,4)
\end{equation}
where ${\textbf k}(2,1) = \hat \rho_{21} \otimes \hat n_{2} - {\textbf I} \hat \rho _{21} \cdot \hat n_{2}$
is the tensor part of $\textbf K$, and the $\textrm{Tr}$ includes an integration of each $\vec r_{i}$ over both surfaces, excluding the cases where all four points are on the same surface.  For all but the case of parallel planes, this expands out into a large number of difficult integrals, and analytic progress seems unlikely.  However, the
numerical integration is straightforward.

\subsection{Higher-order contributions}

This process could be continued to higher $m$.  The analytic integration over $y$ can be done as described above, but the result will be algebraically complex.  Since we will be forced to use numerical integration over the $2m$ two-dimensional surfaces (an integration over a $4m$ dimensional space), we can just as well do the numerical integration over $y$ at the same time.  We will show below that for selected problems, the first two ${\mathcal E}_{m}$ give usably accurate values.   We will also discuss cases involving cylinders, where carrying on to small finite order is insufficient.   
 
\subsection{Comparison with other approaches}

Most routes to the evaluation of the Casimir interaction have made use of a contour integral representation.  The integrand of the contour integral contains a response function that has a pole at every mode frequency.   In the present case, this response function is the relationship between $\vec{j_s}$ and $\vec{j}$ implicitly defined by Eq.(\ref{BD1}).

Rather than directly solving (\ref{BD1}), BD represent the solution as an
infinite series by repeatedly substituting the equation into itself.  This 
has the advantage that the expression for the Casimir interaction is in the
form of a sum of integrals over known functions.   

Analytic progress can be made on the direct solution of (\ref{BD1}) in cases of high symmetry, where the tensor ${\textbf K}$ has a diagonal representation.  This has made possible the evaluations for the interaction of two spheres \cite{EGJK,KK},
sphere and plane \cite{KK,emig,Maia Neto}, cylinder and plane \cite{Emig_cylinder}, and parallel 
cylinders \cite{Rahi}. However, there is little further that can be done
by this approach, since even for the prolate ellipsoid it is unknown how to
separate Maxwell's equations \cite{ellipse}.
Each of these evaluations requires its own mathematical elaboration (expansions in Bessel functions or spherical harmonics), which allow very little generalization.

Subdividing the surfaces of interacting objects into small polygons converts (\ref{BD1}) into a large matrix equation, which can be inverted by standard
algorithms.  This approach has been explored by Johnson and collaborators \cite{johnson}.  
Although this does afford a general route to the evaluation of the interaction between objects of arbitrary shape, its use requires some expertise in 
the art of tessellation of surfaces and inversion of large matrices.   The output of the program is the energy of one configuration (or the stress one
exerts on the other); getting a sense for how the objects interact can only
come from studying the numerical output.

In contrast, the BD expansion describes the interaction in terms of an 
integral (\ref{casimir4}); we have shown that the integral over imaginary
frequency can be done, leading, in the two-point approximation, to the representations
(\ref{int1}), (\ref{int2}), (\ref{int3}),
whose form is simple enough that analytic evaluation is possible for the
kinds of problems that have been worked in the past, and that one can
learn something about the nature of the interaction from it even when
analytic evaluation is not possible.  The two-point approximation is a controlled  
improvement on the PFA: about as easy to use and much more general.

We have shown that the integral over imaginary frequency can also be done
for the four-point interaction, leading to (\ref{fourpt2}), which now has the
form of a geometrical interaction between the objects.  As will be shown 
below, the four-point interaction often is a small correction (indicating
rapid convergence of the BD series).  We discuss the exceptions below.
 
In the forms that we have presented them, the two- and four-point interactions can be readily evaluated numerically.
We have chosen to use a rather basic implementation of the Monte Carlo method
(averaging the integrand over randomly chosen multiplets of points) because
it is easy to code, and the same set of programs could be used for all of the geometries we describe, replacing only the subroutine that describes the
interacting surfaces.  Once it had been set up, the program took about a day (running on a standard desktop computer) to generate the figures we display.

\section{Applications}

To judge the utility of the BD expansion, and in particular the accuracy of the two-point term (\ref{int1}) regarded as a replacement for previous approximations \cite{Derjaguin,Most},  we consider some examples.

\subsection{Parallel planes}

Problems in which one of the surfaces is a plane are simplified because the trace of the product of tensors ${\bf K}$ appearing in (\ref{casimir4}) vanishes when two successive points are on the same plane.  For parallel planes this implies that the integration points must alternately be on one surface or the other.  BD discuss this problem and show that the successive contributions differ only in a factor of $1/m^{4}$, explaining why the two-point contribution is less than the exact answer by the factor $\zeta(4) = \pi^{4}/90 = 1.0824$.

\subsection{Plane or sphere facing arbitrary surface}

Eq. ({\ref{int1}) simplifies quite a bit when surface $B$ is a plane facing surface $A$ whose height relative to $B$ is given by a function $z(x,y)$.   The integration over the plane can be done to give
\begin{equation}
\label{toPFA}
{\mathcal E}_{2} =-\frac{ \hbar c}{8\pi^{2}}\int d\vec S_{B}
\cdot \frac {\hat z}{z^{3}} =-\frac{ \hbar c}{8\pi^{2}}\int \frac {dx dy}{z^{3}}. 
\end{equation}
Thus for this problem, the first term of the BD expansion (up to the factor of $\zeta(4)=\pi^{4}/90=1.0824$) derives the PFA; further corrections come from the higher order terms.  Our earlier results for the interaction of the plane with the sphere, (\ref{sphere_plane}), or the cylinder, (\ref{cylinder_plane}), can be reproduced as a superposition of the interactions with front and back surfaces computed with the help of Eq.(\ref{toPFA}).  

A similar reduction of the integral over both surfaces to an integral over only one can be effected
for the case of a sphere of radius $A$ and an arbitrary surface $R(\theta,\phi)$ in the form
\begin{equation}
\label{toPFA2}
{\mathcal E}_{2} =-\frac{A \hbar c}{4\pi^{2}}\int d S_{A}
\frac {5R^{2}-A^{2}}{(R^{2}-A^{2})^{3}}
\end{equation}

\subsection{Sphere and plane}

Emig \cite{emig}, Kenneth and Klich \cite{KK} and Maia Neto, Lambrecht and Reynaud \cite{Maia Neto} have evaluated the interaction energy for a sphere of radius $A$ with center at distance $D$ from a plane.

Figure 1 compares the results of Emig and those of Kenneth and Klich
to the sum of the two- and four-point terms, in each
case dividing by the
two-point result (\ref{sphere_plane}).
The Casimir interaction varies by six orders of magnitude over the range of
the data in this Figure, but the two-point approximation captures all but $10\%$ of this. 
The four-point term accounts for most of the discrepancy.
\begin{figure}
\includegraphics[width=1.0\columnwidth,keepaspectratio]{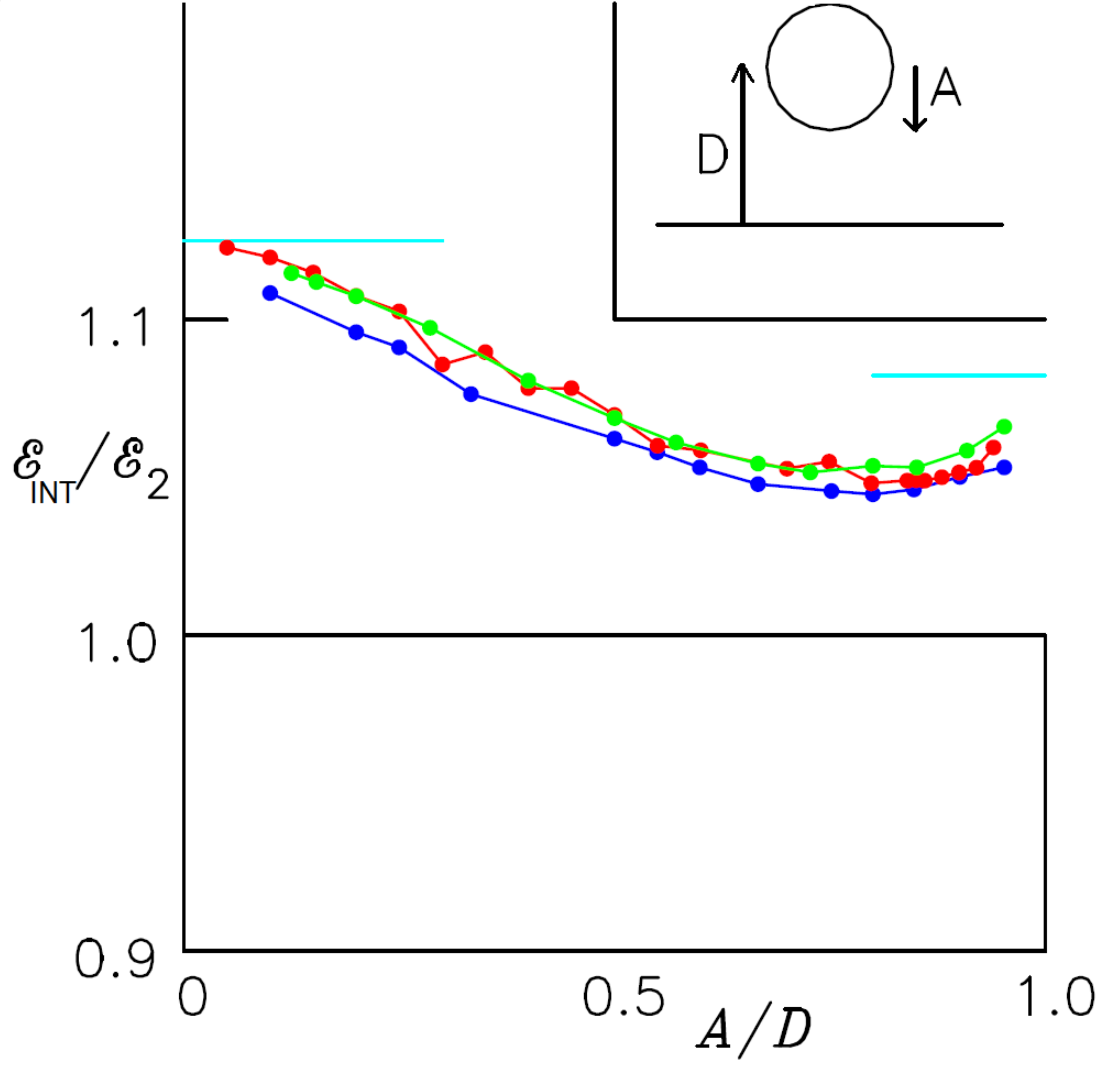}
\caption{(Color online) Casimir interaction energy for a sphere of radius $A$ with center at distance $D$ from
a plane (sketched in the inset).  Green line: Results from Ref. \cite{KK}. Red line: Results from Ref. \cite{emig}.  Blue line: sum of two- and four-point approximations.   These have been divided by the two-point result, Eq.(\ref{sphere_plane}), which removes most of the dependence on $D$ (the straight black line represents the two-point approximation itself).
At the right  we have indicated the \textit{short distance} asymptotic value ${\mathcal E}_{int}/{\mathcal E}_{2} = \pi^{4}/90 = 1.0824$ due to Casimir \cite{Casimir} as discussed in Section IVB.  At the left we indicate the \textit{large distance }asymptotic value ${\mathcal E}_{int}/{\cal E}_{2} = 9/8 = 1.125$ due to Boyer \cite{Boyer}.}
\end{figure}

\subsection{Interaction between spheres}

The interaction between identical spheres of radius $A$ and with centers separated by distance $D$ has been studied by Emig et al. \cite{EGJK} and by Kenneth and Klich \cite{KK}.  The two-point approximation is the special case $A = B$ of (\ref{twospheres}):
\begin{equation}
\label{ABcase}
{\mathcal E}_{2} =-\frac{\hbar c}{8\pi D} \Bigg\{ \ln \left (\frac{D^{2}}{D^{2}-4A^{2}
}\right ) +\frac {40 A^{4} - 6 A^{2} D^{2}}{(D^{2}-4 A^{2})^{2}} + \frac{2A^{2}}{D^{2}} \Bigg\}
\end{equation}
The four-point interaction was evaluated by constructing four points $\vec r_{i}$ randomly chosen to be somewhere on either surface.  The cases where all four were assigned to the same surface were discarded; for the rest, the integrand for ${\mathcal E}_{4}$, Eq.(\ref{fourpt2}), was evaluated and averaged over $10^{9}$ attempts for each value of $D$.  

Figure 2 compares ${\mathcal E}_{2} + 
{\mathcal E}_{4}$ to the results of Ref.\cite{KK} (which are nearly identical to those of Ref.\cite{EGJK}).  Since these are strong functions of $D$, we have divided each result by 
 ${\mathcal E}_{2}$. 
\begin{figure}
\includegraphics[width=1.0\columnwidth,keepaspectratio]{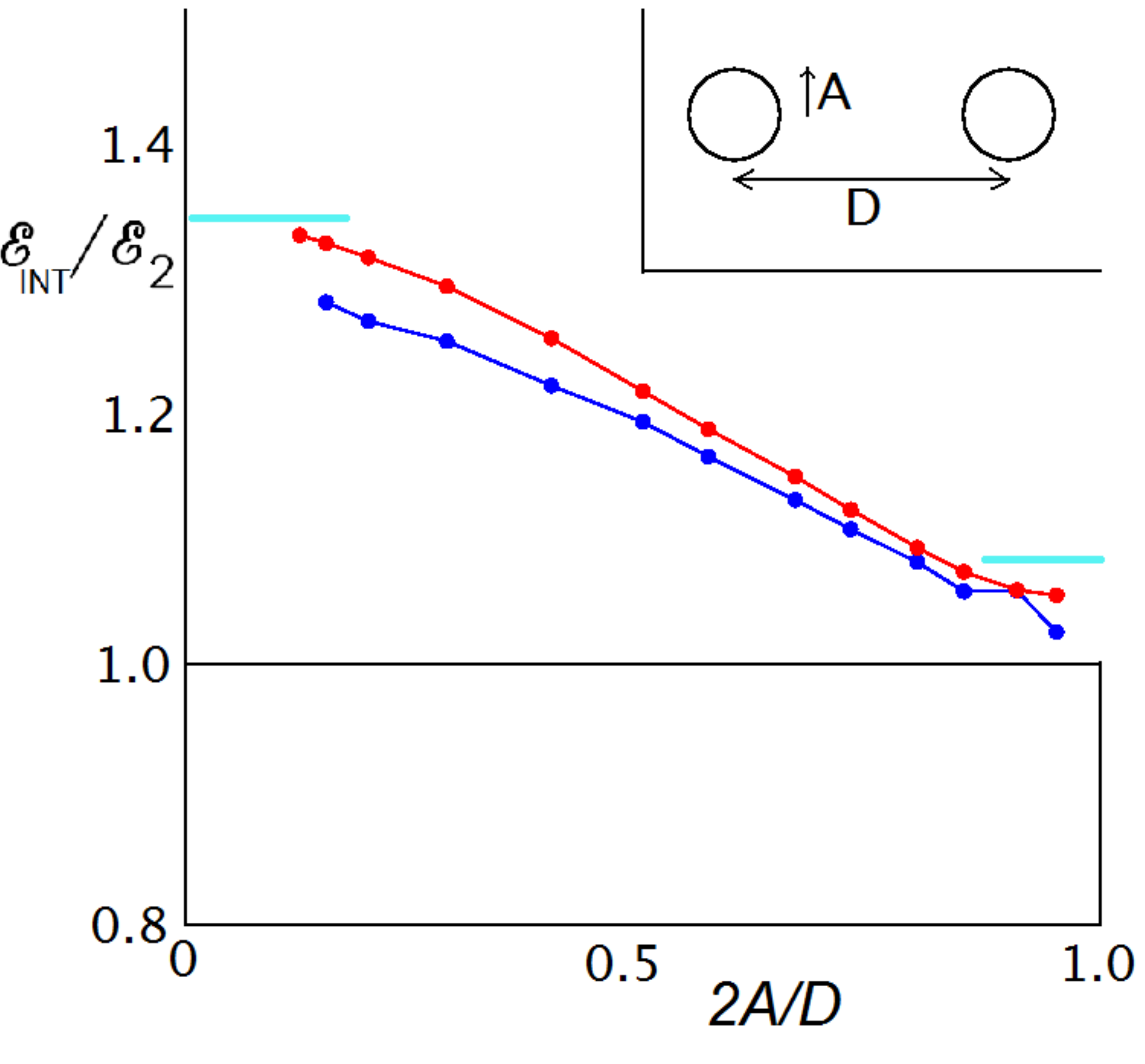}
\caption{(Color online) Casimir interaction energy for spheres of radius $A$ at distance $D$ from each other (sketched in the inset), divided by the two-point interaction energy ${\mathcal E}_{2}$. Black straight line: The two-point interaction Eq.(\ref{ABcase}). Blue line: includes the four-point interaction.  Red line: Results from Ref. \cite{KK}. At the right  we have indicated the asymptotic value ${\mathcal E}_{int}/{\mathcal E}_{2} = \pi^{4}/90 = 1.0824$.  At the left we indicate the asymptotic value ${\mathcal E}_{int}/{\cal E}_{2} = 429/320 = 1.34$.}
\end{figure}

\subsection{Hyperboloid and plane}

The surface $z(x,y) = D + \alpha  \sqrt{A^{2}\alpha^{2} +  x^2+y^2} - \alpha^{2} A$ describes
a cone with a rounded tip, as might be used in an atomic force microscope.
The parameter $\alpha$ is the cotangent of the opening angle of the cone;
the length scale $A$ is the radius of curvature of the tip; and
$D$ is the distance of closest approach to the plane $z = 0$. Eq. (\ref{toPFA}) can be used to find the interaction in the two-point approximation
\begin{equation}
\label{hyperbo}
\mathcal{E}_{2} = - \frac {\hbar c (D+\alpha^{2} A)}{8 \pi \alpha^{2} D^{2}}
\end{equation}
For $D \ll \alpha^{2} A$ this reduces to the small-distance limit of the interaction
between sphere and plane, as given by Eq.(\ref{sphere_plane}) (after making the replacement $ D \rightarrow D + A$ in that equation).  The limit $\alpha \rightarrow 0$ cannot be taken, because the total interaction energy for parallel planes is infinite.  In the $\alpha \rightarrow \infty$ limit the hyperboloid turns into a paraboloid $z(x,y)=D + (x^{2}+y^{2})/2A$ whose interaction energy with the plane in the two-point approximation is $\mathcal{E}_{2}=-\hbar cA/(8\pi D^{2})$.  
 
We evaluated the four-point term as described above, with the results shown in Figure 3.   As previously, we have divided the sum of
$\mathcal{E}_{2} + \mathcal{E}_{4}$ by $\mathcal{E}_2$, Eq.(\ref{hyperbo}), to facilitate comparison.

\begin{figure}
\includegraphics[width=1.0\columnwidth,keepaspectratio]{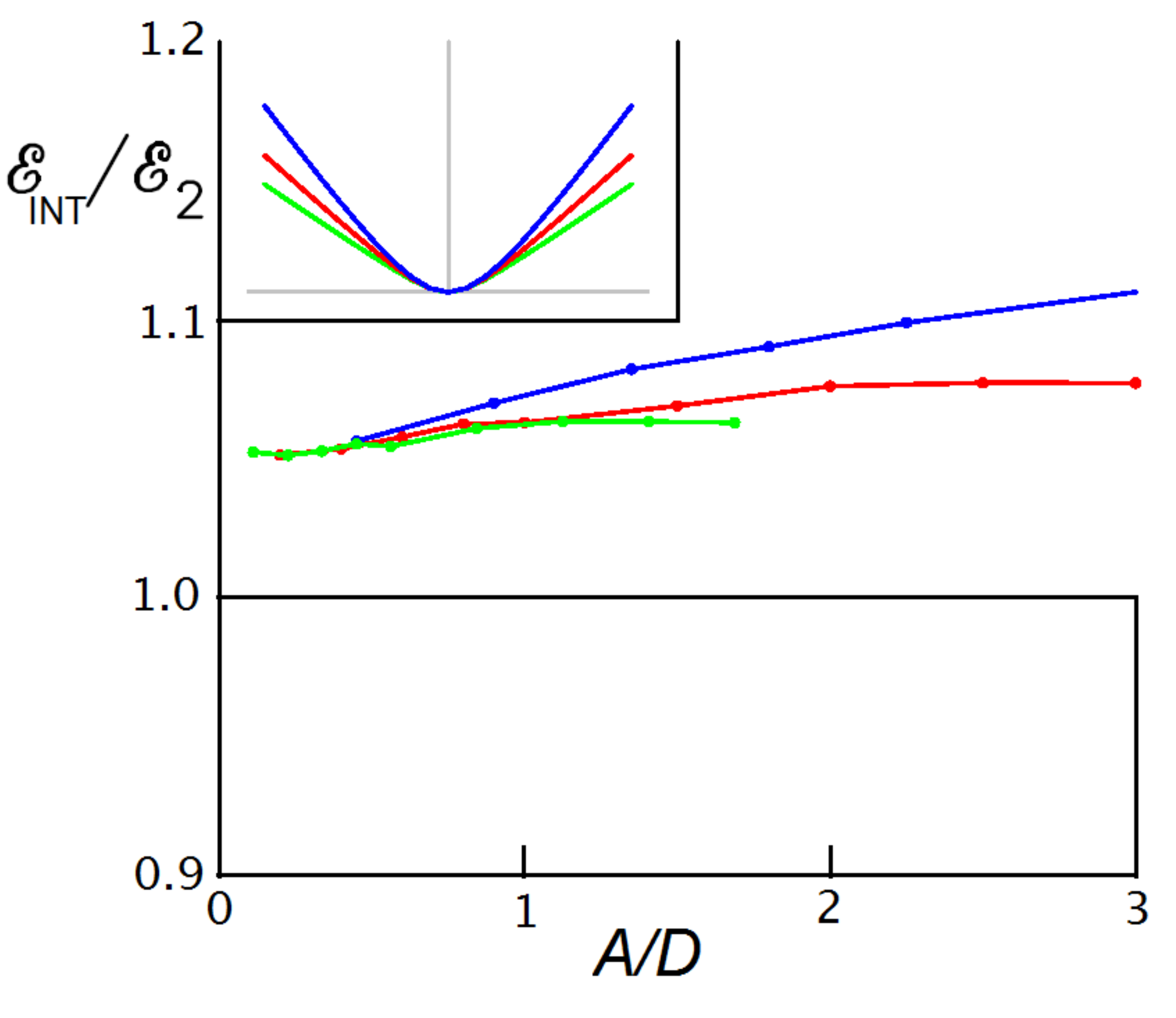}
\caption{(Color online).  Casimir interaction energy between a hyperboloid and a plane (sketched in the inset), as a function of distance of closest approach $D$, for various asymptotic opening angles.
Blue: $\alpha = 1.5$; Red: $\alpha=1.0$; Green: $\alpha = 0.75$. }
\end{figure}

\subsection{Sphere and slab:  illustration of screening}

We can appreciate the need for higher-order terms in the BD expansion by considering the case of a sphere of radius $A$ at distance $D$ from an infinite plane slab of thickness $H$. According to (\ref{int1}) the front surface of the slab will attract the sphere, while the back surface repels it; according to (\ref{int3}) the net interaction is attractive but will become small for small $H$.   In the limit $D \gg A$,
\begin{equation}
\label{sphereslab}
\mathcal{E}_{2}=-C \left ( \frac {1}{D^{4}} - \frac {1}{(D+H)^{4}} \right )
\end{equation}
However, the interaction of the sphere with a conducting plane is attractive, with no need for consideration of what might lie beyond it.  This contradiction arises because the two-point interaction describes an unscreened interaction between the surfaces, yet in the envisioned situation the front surface of the slab will greatly screen the interaction between the sphere and the back surface of the slab. This screening is described by the higher order terms of the BD expansion.

The two terms of (\ref{sphereslab}) arise in (\ref{casimir4}) by the cases where $\vec r_{2}$ is on the the front (F) or back (B) surfaces, respectively, while $\vec r_{1}$ is on the sphere (S). Drawing these as diagrams, sketched in Figure 4 (top)
\begin{figure}
\includegraphics[width=1.0\columnwidth,keepaspectratio]{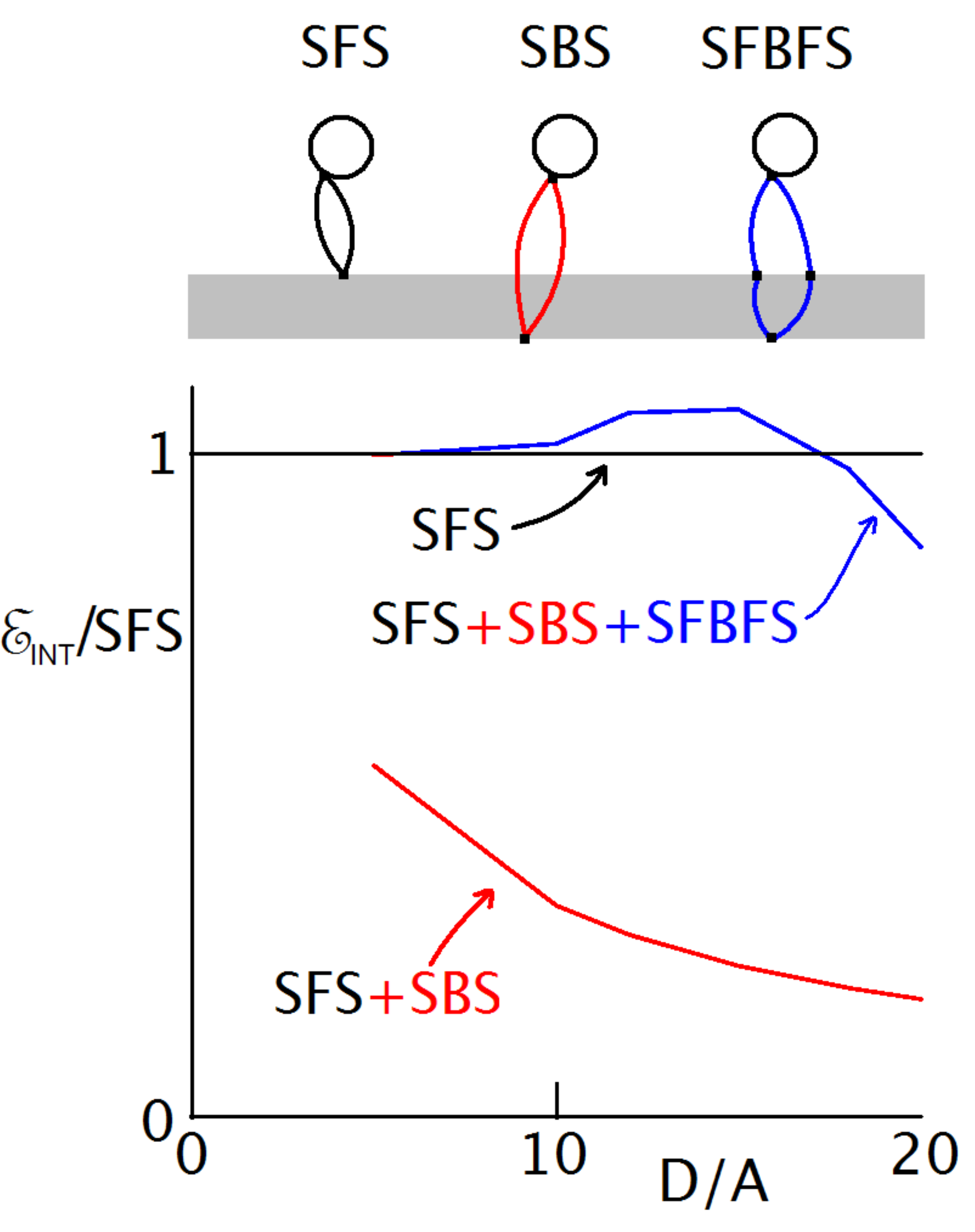}
\caption{(Color online).  Top:  Conducting sphere facing a finite thickness slab (shown in grey);  two-point (SFS and SBS) and four-point (SFBFS) scattering diagrams are sketched.  Bottom:  contribution of these diagrams into the Casimir interaction energy;  the four-point diagram SFBFS essentially cancels the effect of the back surface represented by the SBS diagram thus illustrating screening.  Numerical data employed to produce the graphs are assembled in the Table.}
\end{figure}
the first term is a closed path from sphere to the front surface of the slab and back to the original point on the sphere (represented as SFS).  The unwanted interaction is a closed path from sphere to the back surface and back to the sphere (SBS), crossing the front surface both ways.   The four-point interaction is represented by a large number of diagrams, most of which are corrections to the terms already calculated (such as SSFSS) or identically zero.   We believe that the relevant case is SFBFS, where the interaction between sphere and back surface is modified by crossing the front surface. We calculated this for a unit-radius sphere and unit-thickness slab ($A=1$, $H=1$) by a Monte Carlo integration as described in Section IIID.   To simplify comparison, we have divided all results by $SFS = -C/D^{4}$ (a proxy for the right answer), with the following results:

\begin{table}[h!]
\begin{center}
\begin{tabular}
{l c c c c r}
D & 1 + SBS/SFS & SFBFS/SFS & sum & error & SBFBS/SFS \\
\hline
  5.000 &    0.530 &    0.468 &    0.998 &    0.005 &   -0.274 \\
 10.000 &    0.319 &    0.696 &    1.015 &    0.028 &   -0.501 \\
 12.000 &    0.275 &    0.786 &    1.062 &    0.041 &   -0.562 \\
 15.000 &    0.228 &    0.838 &    1.067 &    0.129 &   -0.661 \\
 18.000 &    0.195 &    0.783 &    0.978 &    0.141 &   -0.622 \\
 20.000 &    0.178 &    0.681 &    0.859 &    0.183 &   -0.925 \\
\end{tabular}
\end{center}
\end{table}

The fourth column shows that SFBFS has substantially canceled SBS for all $D$ (the fifth column is an estimate of the statistical error of the Monte Carlo integration).  We note, however, that another of the four-point diagrams, SBFBS, is large.   We believe it will be canceled at higher order by diagrams such as SFBFBFS.  The data shown in the Table are also displayed in Figure 4. 

The considerations of this paragraph would also be relevant to other systems involving large, thin objects, such as an extremely oblate ellipsoid with a sphere near its axis.  They also explain the success of the PFA (which ignores the effect of the back surface) at short separation distances.   

\subsection{Configurations involving cylinders}

We now assume that in the plane-cylinder geometry the cylinder length $\mathcal{L}$ is effectively infinite, i.e. the interaction energy is strictly extensive in the cylinder length, and focus on the large separation regime $D\gg A$ where 
the two-point prediction, $\mathcal{E}_{2}/\mathcal{L}\simeq -\hbar c A^{2}/D^{4}$, Eq.(\ref{cylinder_plane}), is in qualitative contradiction with the exact result \cite{Emig_cylinder}, $\mathcal{E}_{int}/\mathcal{L}\simeq -\hbar c/(D^{2}\ln(D/a))$. 

\subsubsection{Semi-quantitative argument}

Anomalous interactions arising when long cylinders are involved can be understood via a combination of dimensional analysis and physical reasoning.  In the limit that the length ${\cal L}$ is large compared to other dimensions, we discuss the interaction energy per unit length, which depends on the parameters $\hbar$ and $c$, and on two length scales $D$ and $A$.  Dimensional analysis implies that 
\begin{equation}
\label{dim_analysis cylinder}
\frac{\mathcal{E}_{int}}{\mathcal{L}}=\frac{\hbar c}{D^{2}}f_{c}\left (\frac{A}{D}\right )
\end{equation}
A similar argument applied to the problem of a spherical conductor of radius $A$ placed a distance $D$ away from conductive plane predicts
\begin{equation}
\label{dim_analysis_sphere}
\mathcal{E}_{int}=\frac{\hbar c}{D}f_{s}\left (\frac{A}{D}\right )
\end{equation}    
Here $f_{c,s}(t)$ are dimensionless functions determined by the geometry of the problem.  Both in the case of the cylinder and sphere one must have  $f_{c,s}(0)=0$ because absence of the conductor implies absence of interaction.  However, the character of the approach $f_{c,s}(A/D \rightarrow 0)\rightarrow 0$ is qualitatively different in the two cases.  Introduction of an infinitesimally small spherical conductor is a small perturbation.   Therefore its interaction with the plane as $A/D\rightarrow 0$ must be proportional to the conductor's polarizability, i.e. to its volume, thus implying $f_{s}(t\rightarrow 0)\propto t^{3}$.  This leads to the Casimir-Polder type result (\ref{plane_conductor}).         

Introduction into space of an infinitesimally narrow cylindrical conductor, on the other hand, is a singular perturbation because distribution of the conductive material in space is not compact even if $A$ is infinitesimally small.  The difference between the sphere-plane and cylinder-plane geometries is topological - in the former case any contour in vacuum by continuous deformation can be contracted into a point while this is not true in the latter case.  For the cylinder-plane geometry the topology of space changes at $A=0$.  Therefore the dependence $f_{c}(A/D \rightarrow 0)$ must be non-analytic.  

If topology is indeed the fundamental reason behind the anomalous dependence on separation distance, then non-analytic dependence of the interaction energy on the vicinity to the point of topological change will be relevant to any problem for which the distance between the attracting objects is not the largest length scale.

In order to determine the dependence $f_{c}(A/D)$ we first look at a simpler problem which belongs to the same universality class:  the cavity formed between the two concentric infinitely long cylindrical conductors $\rho>D$ and $\rho< A$, i.e. a coaxial cable.  The reference state here will be chosen to be the absence of the inner conductor, $A=0$.  Then the interaction energy due to insertion of the inner cylinder of radius $A$, directly given by Eq.(\ref{casimirnumber}), must have the functional form (\ref{dim_analysis cylinder}).       

The TEM mode should be omitted as it does not have limiting frequency set by spatial scales of the problem.  The dispersion law of the TE and TM modes is given by 
\begin{equation}
\label{dispersion_law}
\omega^{2}=c^{2}(q^{2}+\kappa^{2})
\end{equation}
where $q$ is the continuous wave number along the cylinder axis while $\kappa$ represents the discrete set of transverse wave numbers corresponding to the cutoff frequencies.  The summation over $\alpha$ in Eq.(\ref{casimirnumber}) now stands for a sum over TM and TE modes with integration over $q$ and summation over all possible $\kappa$.

The spectrum of the TM modes is given by solutions to the transcendental equation
\begin{equation}
\label{trans}
J_{n}(\kappa D)=\frac{J_{n}(\kappa A)N_{n}(\kappa D)}{N_{n}(\kappa A)}
\end{equation}
where $J_{n}$ and $N_{n}$ are the Bessel functions.   The spectrum of the TE modes is given by similar equation except that all the Bessel functions are replaced by their derivatives.

At $A=0$ both the TM and TE equations reduce to $J_{n}(\kappa D)=0$ and $J_{n}'(\kappa D)=0$ which give us the spectrum of the cylindrical cavity of radius $D$.  We are now going to follow solutions to these equations as small $A$ is introduced.  We already know that $A=0$ is the point of non-analyticity of the energy, so we need to look at displacements of the transverse wave numbers which are non-analytic in $A$.  Employing the small argument expansion of the Bessel functions we then infer that only the $n=0$ TM mode produces shifts that are non-analytic in $A$.  So we need to focus on the equation
\begin{equation}
\label{n=0_TM_mode}
J_{0}(x)=\frac{J_{0}(xt)N_{0}(x)}{N_{0}(xt)}, ~~~x=\kappa D, ~~~t=\frac{A}{D}
\end{equation}
This equation is solved iteratively around solutions to the $A=0$ problem $J_{0}(x_{l})=0$ where $l$ labels the roots of the zero-order Bessel function.  The lowest order in $A/D<<1$ solution to (\ref{n=0_TM_mode}) for the displacements $\Delta x_{l}$ due to finite $A$ is  
\begin{equation}
\label{shift}
\Delta x_{l}\approx \frac{\pi N_{0}(x_{l})}{2J_{0}'(x_{l})\ln(A/D)}
\end{equation}
Knowing all the displacements of the transverse wave numbers $\Delta \kappa_{l}=\Delta x_{l}/D$ one can find the change in the energy produced by having $A$ small but finite.  However as far as the $A$-dependence is concerned, this is where we can stop because the entire dependence on $A$ is given by the multiplicative factor of $1/\ln(A/D)$, thus implying $f_{c}(t)\propto 1/\ln t$.  The remaining dependence of the interaction energy per unit length on $D$ follows from dimensional analysis, Eq.(\ref{dim_analysis cylinder}):
\begin{equation}
\label{concentric_cyl_interaction}
\frac{\mathcal{E}_{int}}{\mathcal{L}}\simeq -\frac{\hbar c}{D^{2}\ln(D/A)}
\end{equation}
We see that anomalous dependence on the separation distance is due to the $n=0$ TM mode.  

\subsubsection{Summation of the multiple scattering series}

We will now show that for the cylinder-plane geometry the same anomalous dependence on separation distance (\ref{concentric_cyl_interaction}) follows from a summation of the entire BD series, and that it is a generic feature of the Casimir effect for geometries involving cylinders or wires.  

The geometry is a cylinder of radius $A$ along the $Z$ axis and an infinite conducting plane a distance $D$ from it along $Y$.  We will evaluate (\ref{casimir4}) for the case that ${\vec r}_{2} = (X, D, Z_{2})$ is on the plane and all other ${\vec r}_{i}= (A \cos \phi_{i}, A \sin \phi_{i}, Z_{i})$ are on the cylinder, making approximations appropriate to the limit $D \gg A$.  These paths that make only one visit to the plane are sufficient to explain the result, and are the most important contributions to (\ref{casimir2}) owing to the exponential dependence on path length. Note that in the surface integrals in (\ref{casimir4}) each of the points could be on the plane, and that relabeling so that the point on the plane is always ${\vec r}_{2}$ eliminates the factor of $1/m$ in (\ref{casimir4}). Also note that (\ref{BD2}) vanishes when both ${\vec r}_{1}$ and ${\vec r}_{2}$ are on the plane.  

In cylindrical coordinates, a current $\hat z j_{c}(Z)$ (independent of $\phi$) on the cylinder gives rise to a magnetic field
\begin{eqnarray}
\label{Bout}
\vec B_{out}(\rho,\phi,Z)&=&\hat \phi \int_{-\infty}^\infty  dZ' \nonumber\\
&\times&\int_{-\infty}^\infty \frac {dq}{2\pi}
e^{iq(Z-Z')} kA I_{0}(kA) K_{0}'(k\rho) j_{c}(Z')\nonumber\\
\end{eqnarray}
for $\rho > A$ and by
\begin{eqnarray}
\label{Bin}
\vec B_{in}(\rho,\phi,Z)&=&\hat \phi \int_{-\infty}^\infty  dz'\nonumber\\
&\times&\int_{-\infty}^\infty \frac {dq}{2\pi}
e^{iq(Z-Z')} kA K_{0}(kA) I_{0}'(k\rho) j_{c}(Z')\nonumber\\
\end{eqnarray}
for $\rho < A$, where $I_{0}$ and $K_{0}$ are the modified Bessel functions and $k^{2} = q^{2} + y^{2}$.
The field is discontinuous at the cylinder surface because it is carrying a current; there also is a field due to currents elsewhere (on the cylinder and on the plane).  This nonlocal part of the field is canceled by the local field inside the cylinder, according to the rule
\begin{equation}
\label{cylinderintegraleqn}
j_{c}(Z)= 2 \hat \rho \times \frac {1}{2}(\vec B_{out} + \vec B_{in}
)
\end{equation}
Substituting (\ref{Bout}) and (\ref{Bin}) gives a relationship between the current at different points on the cylinder in the form of an integral similar to (\ref{BD1}) involving a kernel $\textbf {K}_{cc}({\vec r}_1,{\vec r}_2)$.

We can take advantage of the translational symmetry along the cylinder by using a Fourier representation.   Define
\begin{equation}
\label{cylK}
{\textbf K}_{cc}({\vec r}',{\vec r})= \frac {1}{2\pi A}\int e^{iq (Z'-Z)} {\textbf W}_{cc} \frac {dq} {2\pi}
\end{equation}
According to the argument above, the explicit form of ${\textbf W}_{cc}$ is
\begin{eqnarray}
\label{defW}
{{\textbf W}_{cc}}&=& k A[I_{0}(kA) K_{0}'(kA)+I_{0}'(kA) K_{0}(kA) ]
\nonumber
\\
& = & 1-2A k I_{0}'(kA)K_{0}(kA)
\nonumber
\\ 
& \simeq & 
1 - k^{2} A^{2} \ln (1/kA)
\end{eqnarray}
where the small argument approximations for the Bessel functions are taken at the last step;  those are appropriate, since we will only make use of this result for $kD \leq 1$ and $A \ll D$.  By the same argument we will replace $\ln(1/kA)$ by $\ln(D/A)$.
This argument can be repeated for the case that the current on the cylinder varies as $\cos n\phi$.  The effect is to replace the Bessel functions of order $0$ by functions of order $n$.  The important difference 
is that in the small $k$ limit the corresponding ${\textbf W}_{cc}$ is well less than unity.   In the argument to follow, this will
imply that the cylinder does not have resonant response
to this part of the current, so that it can be neglected.

The field (\ref{Bout}) induces a current $\vec j(Z,X) = 2 \hat Y \times \vec B$, and this current gives rise to a field that
can be calculated similarly to (\ref{Bin}) and which in turn requires a screening field on the cylinder.  The corresponding kernels $\textbf K$ can also be given in a Fourier representation, although these also depend on the coordinate $X$ of the plane:
\begin{eqnarray}
\label{defWs}
{\textbf K}(\vec r_{2}, \vec r_{3}) &=& \frac {1}{2\pi A}\int
e^{iq(Z_{2}-Z_{3})} {\textbf W}_{pc}(X) \frac{dq}{2\pi}\nonumber\\
{\textbf K}(\vec r_{1},\vec r_{2}) &=& \int e^{iq(Z_{1}-Z_{2})} {\textbf W}_{cp}(X)\frac{dq}{2\pi}
\end{eqnarray}
The explicit forms of the response functions are
\begin{eqnarray}
\label{Ws}
{\textbf W}_{pc}(q,X) &=& -\frac {2ADk}{R} I_{0}(kA) {K_{0}}'(kR)\nonumber\\
{\textbf W}_{cp}(q,X) &=& \frac {k}{\pi} I_{0}'(kA) K_{0}(kR)
\end{eqnarray}
where $R=\sqrt{X^{2}+D^{2}}$.

After substituting this into (\ref{casimir4}), all of the surface integrals except the integration over $X$ can be done immediately, leading to
\begin{eqnarray}
\label{cylinderinteraction}
\frac{\mathcal{E}_{int}}{\mathcal{L}} &=& - \hbar c \int_{0}^{\infty} \frac {dy}{2\pi} \int_{-\infty}^{\infty} \frac {dq}{2\pi} \nonumber\\
&\times&
\int_{-\infty}^{\infty} {\textbf W}_{cp}(X){\textbf W}_{pc}(X)dX
\sum_{m \ge 2} [{\textbf W}_{cc}]^{m-2}
\nonumber\\
&=&- \hbar c \int_{0}^{\infty} \frac {dy}{2\pi} \int_{-\infty}^{\infty} \frac {dq}{2\pi}\nonumber\\
&\times&\int_{-\infty}^{\infty} dX \frac {{\textbf W}_{cp}(X){\textbf W}_{pc}(X)}
{1-{\textbf W}_{cc}}
\end{eqnarray}
Now the significance of the small argument behavior of ${\textbf W}_{cc}$ becomes clear: because it approaches unity, an arbitrarily large number of terms of the expansion must be included.  The cylinder is exhibiting a resonant response to the presence of the plane.

Substituting the explicit forms for the response functions, the integrals can be done to reproduce the result of Ref.\cite{Emig_cylinder}:
\begin{equation}
\label{cylinderresult}
\frac{\mathcal{E}_{int}}{\mathcal{L}} \approx -\frac {\hbar c}{16\pi D^{2} \ln(D/A)}
\end{equation}
valid for $D \gg A$.

From this discussion it is clear that a similar anomalous Casimir interaction will occur for any geometry involving a cylinder or wire whose length significantly exceeds its radius of curvature.

\section{Effects of finite temperature}

Ideas described up to this point can be extended to the case of finite temperature when the formula for the Casimir energy (\ref{casimir1}) is replaced by an expression for the Casimir free energy \cite{BD}:
\begin{eqnarray}
\label{Casimir_free_energy}
\mathcal{F} &=&  \frac{\hbar c}{\pi} \int_0^{\infty} \left (\Phi(iy) - \Phi(i\infty)\right ) dy\nonumber\\
&+& 2T\int_{0}^{\infty}\frac{dy}{y}[\Phi(iy)-\Phi(+i0)]g(y)
\end{eqnarray}
The temperature effects are accumulated in the second integral and the temperature $T$ is also present in the sawtooth function
\begin{equation}
\label{sawtooth}
g(y)=\frac{1}{2}-\frac{\hbar c}{2\pi T}y+\sum_{n=1}^{\infty}\Theta\left (y-\frac{2\pi nT}{\hbar c}\right )
\end{equation} 
where $\Theta(x)$ is the step function.  Similar to the zero-temperature case, Eqs.(\ref{casimir2}) and (\ref{casimir3}) can be substituted into Eq.(\ref{Casimir_free_energy}) so that the expression for the Casimir free energy of interaction reduces to $\mathcal{F}_{int}=\sum_{m\geqslant1}\mathcal{F}_{2m}$ with $\mathcal{F}_{2m}$ standing for the $2m$-point contribution.  Since at $T=0$ and many geometries the two-point contribution captures the bulk of the effect, we expect the same to hold at finite temperature; some of the evidence to that is given below. In what follows we limit ourselves to the two-point contribution $\mathcal{F}_{2}$ into the Casimir free energy of interaction of conductors $A$ and $B$.  Combining Eqs.(\ref{2_point}),(\ref{Casimir_free_energy}) and (\ref{sawtooth}) we find
\begin{eqnarray}
\label{2_point_fenergy }
\mathcal{F}_{2}&=&\frac{T}{16\pi^{2}}\int \frac{(\vec\rho \cdot d\vec S_{A})(\vec\rho \cdot d\vec S_{B})}{\rho}\nonumber\\
&\times &\frac{d}{d\rho}\Bigg\{\frac{1}{\rho^{2}}\frac{d}{d\rho}\left (\frac{1}{\rho}\coth\frac{2\pi \rho T}{\hbar c}\right )\Bigg\}
\end{eqnarray}
Compared to the zero-temperature result (\ref{int1}), the novel feature of Eq.(\ref{2_point_fenergy }) is the presence of cross-over scale
\begin{equation}
\label{Casimir_length}
\lambda=\frac{\hbar c}{2\pi T},
\end{equation}   
the well-known Casimir length, that defines two regimes of separation of the conductors: 

(i) For $\rho \ll \lambda $ thermal fluctuations are negligible and the zero-point motion dominates the physics.  This is the regime of Casimir/Casimir-Polder interactions.	  
At $T=0$ when $\lambda =\infty$ Eq.(\ref{2_point_fenergy }) reduces to the zero-temperature result (\ref{int1}).  At $T=300K$ we find $\lambda=1\mu m$, thus implying that even at room temperature quantum mechanics dominates Casimir interaction of sub micron separated conductors.  

(ii)  For $\rho \gg \lambda$ thermal fluctuations dominate the physics and zero-point motion is negligible.  This is the regime of Van der Waals interactions.  In the high-temperature regime the Planck's constant drops out of Eq.(\ref{2_point_fenergy }) which acquires purely classical form 
\begin{equation}
\label{fenergy_highT}
\mathcal{F}_{2}= \frac{T}{4\pi^{2}}\int \frac{(\vec \rho \cdot d\vec S_{A})(\vec \rho \cdot d\vec S_{B})}{\rho^{6}}
\end{equation}
As in the zero-temperature case, this can be interpreted as a superposition of vector interactions of elementary surface elements;  in contrast to the $T=0$ situation the interaction (\ref{fenergy_highT}) is more long-ranged.  Application of the Gauss's theorem once and then once again gives two further representations of the result
\begin{equation}
\label{fenergy_highT2}
\mathcal{F}_{2}=-\frac{T}{2\pi^{2}}\int \frac{(\vec \rho \cdot d\vec S_{B})dv_{A}}{\rho^{6}}
\end{equation}  

\begin{equation}
\label{fenergy_highT3}
\mathcal{F}_{2}=-\frac{3T}{2\pi^{2}}\int \frac{dv_{A}dv_{B}}{\rho^{6}}
\end{equation}
which are high-temperature counterparts of Eqs.(\ref{int2}) and (\ref{int3}), respectively.  Eqs.(\ref{fenergy_highT}-\ref{fenergy_highT3}) have the form $\mathcal{F}_{2}=-T\mathcal{S}$ where $\mathcal{S}$ is the Casimir interaction entropy thus implying that in the high-temperature regime the Casimir effect has purely entropic origin.  Eq.(\ref{fenergy_highT3}) also tells us that $\mathcal{S}>0$, i.e. the interaction is attractive (to this order of approximation) for any pair of conductors.  

For simple geometries (plane-plane, plane-sphere, sphere-sphere, and plane-cylinder) it is possible to derive closed form high-temperature counterparts of the results of Section IIIA.  For example, for two conductors separated by a large distance $D$ we find
\begin{equation}
\label{2conductors_highT}
\mathcal{F}_{2}= - \frac{3Tv_{A}v_{B}}{2\pi^{2}D^{6}}
\end{equation}
which captures the functional dependence of the exact result \cite{BG} and has the same order of magnitude.

Similar to the analysis of the zero-temperature and high-temperature limits the general expression (\ref{2_point_fenergy }) can be transformed via Gauss's theorem to find somewhat cumbersome formulas encompassing Eqs.(\ref{int2}), (\ref{int3}), (\ref{fenergy_highT2}), and (\ref{fenergy_highT3}) as special cases.  The main physics consequence is that interaction of any two conductors is attractive (to this order of approximation) for arbitrary temperature.  For the two-plane geometry the integral (\ref{2_point_fenergy }) can be computed for arbitrary temperature, and the interaction free energy can be given in closed form.  This would reproduce the very accurate result due to  BD \cite{BD} who found it differently by specializing the two-point expression (\ref{2_point}) to the case of the two-plane geometry.   Closed form expressions for the interaction free energy can be also given in the Casimir-Polder limit, i.e. for the interaction between well-separated small particles and between small particle and the plane.  The outcome, which is generalization of Eqs.(\ref{2_separate_conductors}) and (\ref{plane_conductor}) to the case of finite temperature, has the same functional form as the result of complementary calculation \cite{Milton} given in terms of particle polarizabilities.    Other geometries that do not allow explicit analytic calculation of $\mathcal{F}_{2}$ can be studied by numerical evaluation of Eq.(\ref{2_point_fenergy }).  

\section{Acknowledgements}

We thank T. Emig for informative correspondence, for sharing with us his numerical data that were employed in producing Figure 1 and for introducing us to the idea of gradient expansion \cite{Bimonte}.   This work was supported by US AFOSR Grant No. FA9550-11-1-0297.

\end{document}